\documentclass[conference,a4paper]{IEEEtran}
\IEEEoverridecommandlockouts
\usepackage[utf8]{inputenc}
\usepackage[T1]{fontenc}   
\usepackage{cite}
\usepackage{amsmath,amssymb,amsfonts}
\usepackage{algorithmic}
\usepackage{graphicx}
\usepackage{textcomp}
\usepackage{xcolor}
\usepackage{tabularx}
\usepackage{array}
\usepackage{makecell}

\begin{document}

\title{Software Product Line Engineering: Adoption, Tooling and AI-Era Challenges}

\author{\IEEEauthorblockN{Najam Nazar}
\IEEEauthorblockA{\textit{Department of Software Systems and Cybersecurity} \\
\textit{Monash University}\\
Australia \\
najam.nazar@monash.edu \\
ORCID: 0000-0003-2317-2297}
}

\maketitle

\begin{abstract}
Software Product Line Engineering (SPLE) enables systematic reuse across families of related software-intensive systems. Following an explicit review protocol covering five databases and four research questions, this survey synthesises SPLE foundations, lifecycle concepts, adoption models, tooling and AI-era challenges. We compare seven adoption and evaluation models (BAPO, FEF, PuLSE, SIMPLE, COPLIMO, PROMOTE-PL and APPLIES), trace four eras of SPLE research, and analyse AI applications with their limitations and assurance requirements. Open challenges in interoperability, UVL standardisation, SME adoption, clone-and-own migration, variability-aware DevOps and empirical evidence are consolidated into a compact research agenda.
\end{abstract}


\section{Introduction}
The growth of software-intensive products in the automotive, telecommunications, consumer electronics and IoT domains has increased the demand for cost effective, high quality software at industrial scale. Software Product Line Engineering (SPLE) addresses this demand through systematic reuse. Instead of developing products independently, organisations build a shared core asset base from which related product variants are derived~\cite{Northrop:2002,Clements:2001}. The central premise of SPLE is to exploit planned commonality while managing intentional variability across a family of systems.

The foundations of SPLE can be traced to early work on program families and domain engineering, including FODA for feature-oriented domain analysis~\cite{Kang:1990}. Subsequent research expanded the field to include feature modelling, automated feature-model analysis, variability implementation techniques, runtime variability, adoption frameworks, maturity evaluation and economic decision support. More recently, SPLE research has begun to address interoperability through the Universal Variability Language (UVL) and AI-assisted techniques for feature extraction, configuration support and variability management~\cite{Benavides:2025}. Nevertheless, tool ecosystems remain fragmented, organisations continue to struggle to manage variability at scale~\cite{Becker:2024}, and AI-assisted methods raise new questions about correctness, explainability and assurance.

\subsection{Research Questions}\label{subsec:rqs}
This survey is guided by four research questions:

\begin{itemize}
\item \textbf{RQ1.} What are the core concepts, lifecycle phases and variability mechanisms that constitute contemporary SPLE?
\item \textbf{RQ2.} How do the principal SPLE adoption and evaluation models differ in purpose, scope, strengths, limitations and reported empirical validation, and which stages of the adoption lifecycle remain unsupported?
\item \textbf{RQ3.} How has SPLE research evolved since its domain-engineering origins, and how have its dominant challenges shifted across successive eras?
\item \textbf{RQ4.} Which technical, organisational and empirical challenges remain open, and what research directions do they imply for standardisation, migration, adoption by small and medium-sized enterprises (SMEs) and AI-assisted variability management?
\end{itemize}

RQ1 is addressed in Sections~\ref{sec:origins} and~\ref{sec:concepts}, RQ2 in Section~\ref{sec:models}, RQ3 in Section~\ref{sec:history}, and RQ4 in Section~\ref{sec:challenges}. This paper contributes a compact taxonomy of SPLE concepts and lifecycle phases; a structured comparison of seven adoption and evaluation models; a four-era synthesis of the evolution of SPLE research; and a consolidated research agenda spanning scalable feature-model analysis, UVL standardisation, clone-and-own migration, SME adoption, empirical evaluation, variability-aware DevOps and AI assurance.

\section{Review Method}\label{sec:method}
This section documents the protocol used to select and analyse the literature.

\subsection{Search Strategy}\label{subsec:search}
We searched five bibliographic databases that are IEEE Xplore, the ACM Digital Library, Scopus, SpringerLink and Web of Science. The search string combined population and intervention terms as (\textit{``software product line''} OR \textit{``product line engineering''} OR \textit{``software product family''}) AND (\textit{``variability management''} OR \textit{``feature model''} OR \textit{adoption} OR \textit{maturity} OR \textit{``clone-and-own''} OR \textit{``machine learning''} OR \textit{``large language model''}). The search covered publications from 1990, the year FODA was published, to March 2026. AI related queries were additionally restricted to 2015--2026, since earlier work predates the techniques of interest. The automated search was complemented by backward and forward snowballing from the foundational references and from the proceedings of SPLC and VaMoS.

\subsection{Selection Criteria}\label{subsec:criteria}
Studies were included if they were (IC1) peer-reviewed and written in English; (IC2) concerned with SPLE concepts, adoption or evaluation models, variability tooling and standardisation, or AI-assisted variability management; and (IC3) presented a method, model, tool, or empirical result rather than a position statement. Studies were excluded if they were (EC1) not primarily concerned with software product lines; (EC2) shorter than four pages and did not present a substantive contribution; (EC3) duplicate or superseded versions of an included study, in which case the most complete version was retained; or (EC4) unavailable in full text. Foundational books and technical reports were retained as explicit exceptions to IC1 because defining contributions, such as FODA~\cite{Kang:1990}, were published in these forms.

Selection proceeded in three stages: title and abstract screening, full text assessment, and backward and forward snowballing. Because record-level search and screening counts were not preserved, a complete study-selection flow cannot be reconstructed retrospectively. The resulting synthesis should therefore be regarded as a structured, representative survey rather than an exhaustive systematic literature review. The final body of literature comprised the works included in the synthesis, supplemented by foundational references used to establish the historical and conceptual background.

\subsection{Data Extraction and Synthesis}\label{subsec:synthesis}
Each included work was classified according to four attributes that are publication era, contribution type (concept, process or adoption model, tool or notation, empirical study, AI technique), the lifecycle phase or BAPO dimension addressed and the type of evidence reported (none, illustrative example, single case, multiple cases, controlled study). The taxonomy in Sections~\ref{sec:origins} and~\ref{sec:concepts} was derived by grouping contribution-type codes and reconciling them against the lifecycle structure of Pohl et al.~\cite{Pohl:2005}. The agenda in Section~\ref{sec:challenges} was derived bottom-up: limitations and future-work statements were grouped by theme, and a theme became an agenda item when raised independently by three or more studies. The eras in Section~\ref{sec:history} mark the years at which the dominant contribution type shifted.

\subsection{Relation to Existing Surveys}\label{subsec:related}
Benavides et al.\ survey automated analysis of feature models~\cite{Benavides:2010}; Chen and Babar review the evaluation of variability management approaches~\cite{Chen:2011}; Raatikainen et al.\ provide a tertiary study of product line and variability modelling~\cite{Raatikainen:2019}; and Berger et al.\ report on industrial variability adoption~\cite{Berger:2020}. These surveys treat analysis techniques, variability modelling, and industrial practice largely as separate concerns. This survey instead compares seven adoption and evaluation models within a single frame, rather than individually or in pairs. This survey also connects the UVL standardisation agenda with assurance requirements for AI-assisted variability management, a topic not addressed by the earlier surveys.

\subsection{Limitations of the Review}\label{subsec:threats}
The review was conducted by a single author. Consequently, study selection and classification were not independently replicated, inter rater reliability could not be calculated, and no formal quality assessment instrument was applied. In addition, record level search, deduplication and screening counts were not preserved, preventing the reconstruction of a complete study selection flow. The resulting body of literature should therefore be regarded as representative rather than exhaustive. Publication bias, database selection bias, and the limited coverage of industrial and grey literature may also have affected the findings.

\section{SPL Origins and Foundations}\label{sec:origins}
The roots of SPLE lie in domain engineering and planned-reuse research. Parnas's argument for designing program families from the outset provided its conceptual foundation~\cite{Parnas:1976}, and Prieto-D\'{i}az contributed domain analysis as a systematic way to identify reuse opportunities~\cite{PrietoDiaz:1990}. The pivotal methodological contribution was FODA~\cite{Kang:1990}, introduced at the SEI in 1990, which made the feature the primary unit of domain characterisation, introduced the feature diagram for mandatory, optional and grouped features, and used cross-tree constraints to encode dependencies. The Family-Oriented Abstraction, Specification and Translation (FAST) process of Weiss and Lai~\cite{Weiss:1999} separated application family engineering from member engineering, while Clements and Northrop~\cite{Clements:2001} synthesised SEI industrial experience into three activity areas---core asset development, product development and management---that remain the canonical reference for SPL adoption.

\section{Core Concepts and Lifecycle}\label{sec:concepts}

\subsection{Feature Models, Core Assets and Standards}\label{subsec:feature}
Feature models are the lingua franca of SPL research. The basic FODA notation~\cite{Kang:1990} was extended with formal semantics by Schobbens et al.~\cite{Schobbens:2007} and with decoupled variability specifications by the Orthogonal Variability Model (OVM)~\cite{Pohl:2005}. The community has historically struggled with notation fragmentation: FeatureIDE and the Linux Kconfig system, for example, use incompatible formats. The Universal Variability Language (UVL)~\cite{Benavides:2025} represents the most ambitious standardisation effort, offering a textual, tool-agnostic format with formally defined semantics that can serialise feature models independently of a specific environment, and UVLHub~\cite{RomeroOrganvidez:2024} complements it with an open science repository enabling reproducible benchmarking.

Core assets are the reusable building blocks of the platform. A reference architecture (an architecture parameterised by variability) is the central core asset around which components, test suites and documentation templates are organised. Feature-Oriented Software Development~\cite{Apel:2009} and delta-oriented programming~\cite{Schaefer:2010} provide composition mechanisms with stronger modularity properties than preprocessor-based approaches, operationalising variability decisions across the compiletime, linktime and runtime binding modes classified by K\"{a}stner et al.~\cite{Kastner:2008}.

\subsection{SPLE Lifecycle}\label{subsec:lifecycle}
\begin{figure}[!t]
    \centering
    \includegraphics[width=0.82\linewidth]{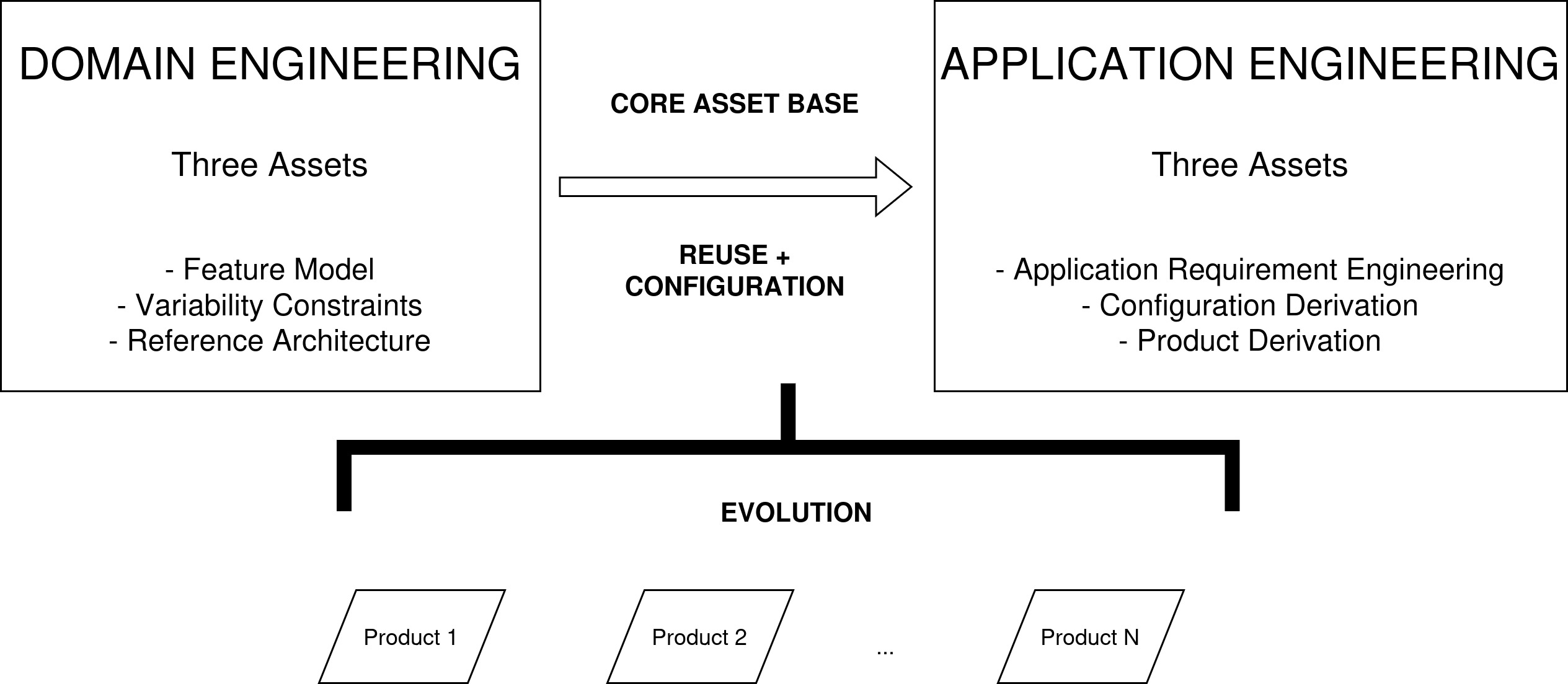}
    \caption{SPL lifecycle: Domain Engineering builds core assets; Application Engineering derives products; evolution spans both.}
    \label{fig:lifecycle}
\end{figure}

The SPL lifecycle, formalised by Pohl et al.~\cite{Pohl:2005}, is organised into two primary activity areas---Domain Engineering and Application Engineering---supported by an orthogonal management area, as shown in Figure~\ref{fig:lifecycle}.

\textit{Domain Engineering} comprises domain requirements engineering (producing the feature model), domain design (the reference architecture) and domain implementation (reusable components, frameworks and test assets); poor variability modelling at this stage propagates complexity throughout the lifecycle, and domain scoping remains under-investigated empirically~\cite{Lindohf:2021}. \textit{Application Engineering} derives individual products from the core assets through application requirements engineering, product-specific adaptation of the reference architecture and product derivation; manual derivation risks defect injection and limits scalability. \textit{Evolution} is operationally critical but under-researched~\cite{Knieke:2022}: extending the feature model, deprecating features or restructuring the architecture ripples across derived products. Adoption and evolution strategies are commonly classed as proactive, reactive or extractive~\cite{Knieke:2022}; the extractive strategy matters in practice because most organisations start from a portfolio of ad hoc variants rather than a clean-slate platform.

\section{SPL Adoption Models}\label{sec:models}
A practically important strand of SPL research addresses how organisations adopt, plan and evaluate the maturity of their product-line practices. Table~\ref{tab:spl_models} compares seven influential models along purpose, scope, strengths, limitations and reported empirical validation; the subsections that follow discuss each in turn.

\begin{table*}[!t]
\caption{Comparison of SPL Adoption and Evaluation Models}
\label{tab:spl_models}
\centering
\scriptsize
\renewcommand{\arraystretch}{1.1}
\setlength{\tabcolsep}{4pt}
\begin{tabularx}{\textwidth}{|>{\raggedright\arraybackslash}p{1.4cm} |>{\raggedright\arraybackslash}p{1.9cm} |>{\raggedright\arraybackslash}X |>{\raggedright\arraybackslash}X |>{\raggedright\arraybackslash}X |>{\raggedright\arraybackslash}X|}
\hline
\textbf{Model} & \textbf{Purpose} & \textbf{Scope / Dimensions} & \textbf{Strengths} & \textbf{Limitations} & \textbf{Reported Validation} \\
\hline

BAPO~\cite{Linden:2007} &
Adoption framework &
Four interdependent dimensions: Business, Architecture, Process, Organisation &
Holistic; makes cross-dimension interdependence explicit; underpins FEF &
Descriptive; lacks an assessment procedure; based mainly on large organisations &
Applied in avionics~\cite{Lindohf:2021} and in an SME case study~\cite{Nazar:2016} \\
\hline

FEF~\cite{Linden:2004,Linden:2007} &
Maturity evaluation &
Five-level scale per BAPO dimension across 13 assessed aspects &
Produces maturity profiles and actionable roadmaps &
Requires substantial tailoring and domain knowledge &
Nine avionics product lines over eleven months at Saab~\cite{Lindohf:2021} \\
\hline

PuLSE~\cite{Bayer:1999} &
Process methodology with maturity scale &
Construction and evolution activities; PuLSE-Eco (scoping), PuLSE-CDA (cost--benefit); four maturity levels &
End-to-end methodology; integrates scoping and economic analysis into the process &
PuLSE-specific maturity levels fit organic adopters poorly~\cite{Lindohf:2021} &
Industrial application in embedded domains; limited recent independent evaluation \\
\hline

SIMPLE~\cite{Clements:2005} &
Management-level cost model &
Cost functions for organisational investment, platform, reuse, product-specific software, maintenance &
Usable before detailed engineering data exist; supports investment framing &
Coarse grained; sensitive to early assumptions; no variability specific granularity &
Hypothetical scenarios and SPL practice patterns; no industrial validation reported \\
\hline

COPLIMO~\cite{Boehm:2004} &
Fine-grained cost and ROI model &
COCOMO~II-based; separates new, reused and adapted product-specific software &
Quantitative ROI estimates; explicit reuse and adaptation factors &
Requires size and effort data unavailable at inception &
Example-based analysis; later independently evaluated and refined; broad industrial calibration limited \\
\hline

PROMOTE-PL~\cite{Kruger:2020} &
Round-trip process model for extractive adoption &
Migration from clone-and-own variants to a platform; incremental scope expansion &
Matches the common industrial starting point; supports incremental commitment &
Automated support for variability mining and effort estimation remains immature &
Derived from industrial practice; longitudinal validation limited \\
\hline

APPLIES~\cite{Rincon:2018} &
Pre-adoption readiness assessment &
Motivation and preparation dimensions; maps onto the BAPO pillars &
Addresses the pre-commitment stage left open by BAPO and FEF &
Qualitative; adoption outcomes following assessment are not yet reported &
Two experts initially; one practitioner, five SPL experts and 14 software practitioners \\
\hline

\end{tabularx}
\end{table*}

\subsection{BAPO and the Family Evaluation Framework}\label{subsec:bapo}\label{subsec:FEF}
The BAPO model~\cite{Linden:2007} defines four interdependent adoption dimensions. Business covers product-line strategy and the funding of platform development independently of individual product revenues; Architecture covers the reference architecture, core assets, feature models and variability mechanisms; Process covers the adaptation of practice to the dual structure of domain and application engineering; and Organisation covers team structure, governance and the cultural transition to platform-oriented engineering. Its key contribution is alignment: exposing feature selection to customers, for example, requires feature models understandable to non-technical stakeholders, which reshapes architecture, process responsibilities and governance. Nazar and Rakotomahefa~\cite{Nazar:2016} extended BAPO-based analysis to SMEs in an ethnographic study of a small Swedish company, finding that SPL-aligned practices had emerged informally but that full adoption was limited by weak Process and Organisation alignment.

The Family Evaluation Framework (FEF)~\cite{Linden:2004,Linden:2007} rates each BAPO dimension on a five-level scale (Independent Development, Standardised Infrastructure, Software Platform, Variant Products, Full Configuration) across 13 aspects, producing a maturity profile rather than a single score. Applying it to nine avionics product lines at the Saab Simulation Centre over eleven months, Lindohf et al.~\cite{Lindohf:2021} found that it supported shared understanding, cross-product-line comparison and actionable roadmaps, but required substantial tailoring and domain knowledge; it is therefore best used as a context-sensitive instrument rather than a checklist.

\subsection{PuLSE, SIMPLE and COPLIMO}\label{subsec:pulse}\label{subsec:simple}
PuLSE~\cite{Bayer:1999} structures adoption around construction and evolution activities, supported by components such as PuLSE-Eco for scoping and PuLSE-CDA for cost--benefit analysis, with four maturity levels (Initial, Full, Controlled, Optimising). Because maturity is defined by the adoption of PuLSE-specific components, organisations with mature but organically developed practices may align poorly with its milestones~\cite{Lindohf:2021}, unlike FEF, which is independent of any process model. SIMPLE~\cite{Clements:2005} defines high-level cost functions for organisational investment, platform development, reuse, product-specific software and maintenance, supporting economic reasoning before engineering data exist, while COPLIMO~\cite{Boehm:2004} offers a more quantitative lifecycle cost and return-on-investment (ROI) model distinguishing new, reused and adapted software. Both inherit a general weakness of cost estimation, since inputs are most uncertain at inception, when the investment decision is made; their value lies in making assumptions explicit and giving technical and managerial stakeholders a shared vocabulary.

\subsection{PROMOTE-PL and APPLIES}\label{subsec:applies}
PROMOTE-PL~\cite{Kruger:2020} is a round-trip process model for the common case in which organisations start from independently developed variants, supporting incremental migration by recovering commonality and variability, establishing an initial platform and evolving it as further variants are incorporated. APPLIES~\cite{Rincon:2018} assesses an organisation's motivation and preparation before adoption. The two are complementary: APPLIES informs the readiness stage, whereas PROMOTE-PL guides the engineering transition, and together they extend the adoption perspective beyond diagnostic maturity assessment.

\section{Historical Evolution of SPL Research}\label{sec:history}
Table~\ref{tab:spl_history} summarises four overlapping eras of SPL research. Across these eras, challenges have shifted from representing variability and justifying platform investment towards scaling analysis, improving interoperability, supporting migration and assuring AI-assisted methods.

\begin{table}[!t]
\caption{Historical Evolution of SPL Research}
\label{tab:spl_history}
\centering
\footnotesize
\setlength{\tabcolsep}{4pt}
\renewcommand{\arraystretch}{1.25}
\begin{tabularx}{\columnwidth}{
|>{\raggedright\arraybackslash}p{0.19\columnwidth}
|>{\raggedright\arraybackslash}X
|>{\raggedright\arraybackslash}X|}
\hline
\textbf{Era} & \textbf{Representative Work} & \textbf{Dominant Challenges} \\
\hline

\textbf{1990--2002}\newline Foundations and planned reuse &
FODA~\cite{Kang:1990}; FAST~\cite{Weiss:1999}; Clements and Northrop~\cite{Clements:2001} &
Formalising variability; limited integrated tooling; justifying upfront investment \\
\hline

\textbf{2003--2012}\newline Formal analysis and adoption frameworks &
BAPO~\cite{Linden:2007}; automated analysis~\cite{Benavides:2010}; FOSD~\cite{Apel:2009}; delta-oriented programming~\cite{Schaefer:2010}; dynamic SPLs~\cite{Hallsteinsen:2008}; PuLSE, SIMPLE, COPLIMO~\cite{Bayer:1999,Clements:2005,Boehm:2004} &
Scalability of analysis; tool interoperability; feature-model evolution; fragmented methods \\
\hline

\textbf{2013--2019}\newline Tool maturation and industrial evaluation &
FeatureIDE~\cite{Thum:2014}; SME-oriented BAPO study~\cite{Nazar:2016}; APPLIES~\cite{Rincon:2018}; adoption studies~\cite{Berger:2020} &
Feature-model maintenance at scale; organisational readiness; migration cost; thin cross-domain evidence \\
\hline

\textbf{2020--present}\newline Standard\-isation and AI-assisted variability &
PROMOTE-PL~\cite{Kruger:2020}; FEF at scale~\cite{Lindohf:2021}; AI-assisted re-engineering~\cite{Acher:2023}; UVLHub~\cite{RomeroOrganvidez:2024}; UVL~\cite{Benavides:2025} &
LLM correctness; explainability; regulatory compliance; benchmark availability \\
\hline

\end{tabularx}
\end{table}

The Foundational Era centred on representing variability and justifying upfront platform investment. The Consolidation Era added automated feature-model analysis, alternative implementation mechanisms such as FOSD and delta-oriented programming, runtime variability through dynamic SPLs~\cite{Hallsteinsen:2008}, and the adoption, maturity and economic models of Section~\ref{sec:models}. The Industrialisation Era emphasised tool support, notably FeatureIDE~\cite{Thum:2014}, and the organisational conditions for adoption~\cite{Nazar:2016,Rincon:2018}. The current era focuses on interoperability through UVL and UVLHub, incremental migration from clone-and-own portfolios, stronger empirical grounding~\cite{Lindohf:2021,Berger:2020}, and the correctness and trust questions raised by AI-assisted methods (Section~\ref{subsec:ai}).

\section{Open Challenges and Future Research}\label{sec:challenges}
Several technical, organisational and empirical challenges remain unresolved. Each challenge below was raised independently by three or more studies in the retained pool, as described in Section~\ref{subsec:synthesis}.

\subsection{Scalable and Trustworthy Feature-Model Analysis}
Industrial feature models such as Linux Kconfig may contain thousands of features and complex constraints, making satisfiability checking, dead-feature detection, configuration counting and impact analysis difficult. BDD-, SAT- and sampling-based techniques have improved scalability for selected tasks, and uniform sampling shows promise for highly configurable systems~\cite{Heradio:2022}, but scalable exact reasoning with strong correctness guarantees remains challenging for very large, constraint-rich product lines. Future work should investigate hybrid methods that combine exact reasoning, approximation and sampling while making the trade-off between scalability and soundness explicit, particularly in continuous engineering settings where models evolve frequently.

\subsection{Interoperability, UVL Standardisation and Open Infrastructure}
The tooling landscape remains fragmented across FeatureIDE, pure::variants, Kconfig and SPLOT; import and export mechanisms exist, but differences in modelling concepts, constraint languages and semantics hinder direct exchange. UVL has emerged as a promising human-readable interchange language~\cite{Benavides:2025}, and UVLHub provides open infrastructure for reproducible studies~\cite{RomeroOrganvidez:2024}. Open problems include preserving constraints during conversion, validating semantic equivalence after import and export, representing industrial variability concepts that do not map onto standard constructs, and integrating open repositories with benchmarking, CI pipelines and versioned datasets.

\subsection{Migration from Clone-and-Own to Platform-Based Reuse}
Many organisations maintain portfolios of cloned or independently evolved variants rather than a planned platform. PROMOTE-PL provides a process for adopting product lines from existing variants~\cite{Kruger:2020}, and research on product-line evolution highlights the long-term architectural consequences of platform change~\cite{Knieke:2022}, yet automated support for mining variability, identifying reusable assets and estimating migration effort remains limited. Extractive adoption would benefit from automated detection of common and variable functionality across clones, traceability between recovered features and code, and decision support for which assets to stabilise first.

\subsection{Lightweight Adoption Pathways for SMEs}
BAPO and FEF were developed primarily in large industrial contexts~\cite{Linden:2007,Lindohf:2021}, whereas smaller organisations face limited resources, flatter structures, shorter planning horizons and narrower product-family scopes. SMEs may exhibit informal SPL-aligned practices without explicit alignment across the BAPO dimensions~\cite{Nazar:2016}, and APPLIES highlights the need to assess motivation and preparation before adoption~\cite{Rincon:2018}. Lightweight frameworks should support incremental adoption, limited upfront investment and staged alignment, helping firms decide when systematic reuse is justified, which assets should become core assets, and how much process formalisation suits their scale.

\subsection{Empirical and Economic Evaluation}
Longitudinal evidence on the full economic lifecycle of SPLE remains limited. Many widely cited benefit claims rest on early industrial case studies, while recent empirical work focuses on qualitative adoption challenges, readiness and maturity~\cite{Lindohf:2021,Berger:2020}, leaving weak evidence on return on investment, maintenance and platform-evolution cost. Future studies should evaluate overlooked costs, including feature model maintenance, platform governance, test infrastructure and organisational coordination, against reuse benefits.% SIMPLE and COPLIMO remain valuable but require calibration against contemporary industrial data.

\subsection{AI-Assisted Variability Management and Assurance}\label{subsec:ai}
AI techniques now touch most phases of the SPLE lifecycle. Table~\ref{tab:ai_spl} organises current applications by the technique employed, the dominant limitation reported in the literature, and the assurance mechanism required before industrial use in a regulated setting.

\begin{table}[!t]
\caption{AI Applications in SPLE: Limitations and Assurance Requirements}
\label{tab:ai_spl}
\centering
\scriptsize
\setlength{\tabcolsep}{3pt}
\renewcommand{\arraystretch}{1.2}
\begin{tabularx}{\columnwidth}{
|>{\raggedright\arraybackslash}p{0.24\columnwidth}
|>{\raggedright\arraybackslash}X
|>{\raggedright\arraybackslash}X|}
\hline
\textbf{Application (technique)} & \textbf{Dominant Limitation} & \textbf{Assurance Requirement} \\
\hline
Feature extraction from requirements(NLP, LLM prompting)~\cite{SreeKumar:2018} &
Natural language ambiguity; features proposed that are absent from the source &
Traceability from each feature to its source artefact; expert confirmation \\
\hline
Reengineering variants into a product line (generative AI)~\cite{Acher:2023} &
Stochastic outputs; degradation on large or complex artefacts &
Behavioural equivalence testing of derived products against the originals \\
\hline
Feature-model validation and repair (LLM-based)~\cite{Le:2026} &
No soundness guarantee independent of a formal reasoner &
Cross-checking against SAT- or BDD-based analysis \\
\hline
Configuration recommendation (recommender systems)~\cite{Pereira:2018} &
Bias towards frequent configurations; cold start &
Constraint satisfaction checked against the feature model; explanation \\
\hline
Performance prediction (regression, transfer learning)~\cite{Guo:2013} &
Sampling bias; limited generalisation across versions &
Reported uncertainty bounds; revalidation after platform evolution \\
\hline
Product selection and optimisation (search-based, RL)~\cite{LopezHerrejon:2014} &
Local optima; sensitivity to objective specification &
Validity checking against the feature model; transparent trade-off reporting \\
\hline
\end{tabularx}
\end{table}

Two patterns are visible. First, reported benefits primarily concern effort reduction rather than demonstrated correctness, AI shortens the path to a candidate artefact without establishing that it is right. Second, every assurance requirement reduces to one of three mechanisms---checking output against a formal reasoner, tracing it to a source artefact, or quantifying its uncertainty---and none is currently integrated into mainstream variability management tooling.

This gap matters most in safety-critical domains such as automotive, avionics and medical devices, where configuration decisions feed certification arguments under standards such as ISO~26262, DO-178C and IEC~62304. These require requirement-to-implementation traceability and evidence of verification coverage, which an unverified generated artefact cannot provide. Research is needed on pipelines that verify AI-generated feature models and configurations against formal constraints, on explanations usable by domain engineers rather than machine-learning specialists, and on shared benchmarks, plausibly built on UVLHub~\cite{RomeroOrganvidez:2024}, so that AI-assisted techniques are compared on common data rather than individual case studies.

\subsection{Variability-Aware Testing, DevOps and CI/CD}
Testing and continuous delivery remain difficult because faults, builds and deployments may be configuration specific and product spaces grow exponentially with the number of features. Sampling, combinatorial interaction testing, lifted verification and variability aware regression testing reduce this burden but depend on accurate feature models, feature-aware test suites and automated derivation infrastructure~\cite{Heradio:2022}. Future studies should examine how CI/CD pipelines can identify the configurations affected by a change, select representative products for testing, and maintain traceability between features, code, tests and deployment artefacts; PROMOTE-PL offers process-level guidance~\cite{Kruger:2020}, but its operationalisation through automated build selection remains open.

\section{Conclusion}\label{sec:conc}
Following the protocol in Section~\ref{sec:method}, this survey has traced the evolution of SPLE from its domain engineering foundations to the emerging use of AI-assisted techniques. It answers RQ1 and RQ2 with an account of SPLE concepts and lifecycle phases and a structured comparison of seven adoption and evaluation models, RQ3 with a four era synthesis, and RQ4 with a consolidated set of open challenges. Although SPLE is technically mature, the field remains fragmented across tools, modelling formats, adoption practices and empirical evidence; the priorities in Section~\ref{sec:challenges} define a compact research agenda for software engineering and ICT researchers.

\bibliographystyle{IEEEtran}
\bibliography{references}

\end{document}